# Magnetically dependent plasmon drag in permalloy structures

**Mohammad Shahabuddin\*,1, David W. Keene1, Maxim Durach2, Vladimir S. Posvyanskii3, Vadim A. Atsarkin4, Natalia Noginova1**

*1Center for Materials Research*
*Norfolk State University, Norfolk, Virginia 23504, United States*
*2Georgia Southern University, Statesboro, Georgia 30458, United States*
*3Semenov Federal Research Center of Chemical Physics of RAS, Moscow, 119991, Russia*
*4Kotel'nikov Institute of Radio Engineering and Electronics of RAS, Moscow, 125009, Russia*

*Corresponding Author: m.shahabuddin@spartans.nsu.edu*
*\* Tel: 757-823-2298*

**Abstract.** Significant photovoltages are observed in permalloy grating-like structures in response to pulsed laser light illumination. Electric signals are enhanced at plasmon resonance conditions and show a clear dependence on the magnetic field with a characteristic hysteresis. Estimations show that the effect could not be explained solely by the laser-induced heating. Alternative mechanisms are discussed.

Coupling of plasmonic and magnetic effects can bring many technological possibilities and applications such as compact magneto-optical devices, fast magnetization switching and nonreciprocal plasmonics[1]. The experiments in this direction[2-7] involve core-shell nanoparticles having a magnetic core (Co, Fe-Co, iron oxide) and a plasmonic shell (Au), or nanocomposite films and heterostructures of plasmonic and magnetic materials. Significant enhancement of magneto-optical activity and an increase in photomagnetization under illumination with circular polarized light at plasmon resonance conditions clearly indicate participation of surface plasmons in a spin angular momentum (SAM) exchange between light and matter. The effects are explained with strongly enhanced electric fields at resonance conditions, and also discussed in terms of the enhanced inverse Faraday effect[8].

As shown in[9-11] surface plasmon polaritons (SPPs) propagating at the interface of a dielectric and a metal have SAM associated with the rotation of the optical electric field polarization in the skin layer. Transfer of SAM from plasmons to a material can result in substantial spin polarization. In our work we look for the signature of this process, observing photocurrents generated in response to surface plasmon polariton (SPP) excitation in systems which combine plasmonic and magnetic properties. Giant enhancement of photocurrents in gold and silver structures observed at SPP resonance (plasmon drag effect[12-18]) is commonly explained by linear momentum exchange between light and matter and described with the electromagnetic hydrodynamic momentum loss (EHML) approach[19,20]. However, other factors (surface charges, etc.) can contribute to photoinduced electric effects in plasmonic metal[21, 22] as well. If propagating plasmons affect spin polarization, the overall picture of the photoinduced electric effects might be very different[23] from that in a purely non-magnetic case, bringing another dimension to this research area and additional possibilities to control plasmonic excitations and electric effects associated with plasmons.

Our experimental structures are permalloy (Ni-Fe alloy with 80% of Ni and 20% of Fe) thin films with one-dimensional (1D) profile modulation based on the substrates derived from commercially available DVD-R discs. Similar systems based on BluRay (BR) substrates having a different profile modulation are fabricated as well for comparison purposes. When covered with plasmonic metal both geometries provide conditions for SPP excitation via grating configuration[24].

The fabrication starts with obtaining polycarbonate grating substrates from disassembling commercial DVD-R or BluRay (BR) discs by carefully taking out the polymer, plastic, silver and protective coating layers. Then, permalloy (Py) with a thickness $d = 40$ nm is deposited on the prepared and precut DVD, BR, and glass substrates using e-beam evaporation. The thickness of the film is independently tested with a profilometer by measuring films simultaneously deposited on glass substrates. Atomic force microscopy (AFM) confirms the profile modulation parameters, the periodicity $p = 740$ nm and modulation height $h =$60-80 nm in Py/DVD, and $p = 320$ nm and $h =$ 25-30 nm in Py/BR structures respectively, see Figs. 1(a-d) for the schematics and AFM images. For optical characterization, the samples are illuminated with p-polarization with the grooves oriented perpendicular to the incidence plane (Fig. 1(e)). The reflectivity spectra show well-pronounced dips, which correspond to conditions for grating-coupled surface plasmon resonances[25].

$$\mathbf{k_{spp}} = m\mathbf{G} + \mathbf{k_x} \qquad (1)$$

where $k_x = k_0 \sin\theta$ is the projection of the optical wave-vector $k_0$ onto the film plane, $\theta$ is the angle of incidence, $k_{spp}$ is the SPP wave-vector, $G = 2\pi/p$ is the grating vector, and $m$ is an integer. The dips shown in Fig. 1 (f) for Py/DVD correspond to $m = 1$ while those in Py/BR, Fig. 1 (g) can be fitted with $m = -1$ correspondingly. The $\omega(k)$ dispersion curve derived from spectral positions of the dips at various angles at both structures is shown in Fig. 1 (h). As one can see, it is different form the photon line (dashed trace) and fits well the estimation (red solid curve) calculated from the relationship[25]

$$k_{SPP} = \frac{\omega}{c}\sqrt{\frac{\varepsilon_m \varepsilon_d}{\varepsilon_m + \varepsilon_d}} \quad , \qquad (2)$$

where $\omega$ is the frequency, $c$ is the speed of light, $\varepsilon_d = 1$ is the permittivity of air, and $\varepsilon_m$ ($\omega$) is the real part of the permalloy permittivity from Ref.[26]. Q-factors of the resonances estimated from the dip width are relatively low, in the range of 3-10 depending on the wavelength; this can be expected due to the high imaginary part of the permittivity[26].

Permalloy is a soft ferromagnetic with a very high magnetic susceptibility (up to 90000[27, 28]). As shown in Refs[29,30], 1D-profile modulation of a magnetic film with submicron periodicity can produce uniaxial magnetic anisotropy with in-plane anisotropy axis along the grooves. Magnetic properties of our structures are characterized with the ferromagnetic resonance (FMR) method, which confirms the presence of in-plane uniaxial magnetic anisotropy with the anisotropy field $H_a$ up to 90 Oe in Py/DVD samples[31], and of lower values (of ~ 50 Oe) in Py/BR samples.

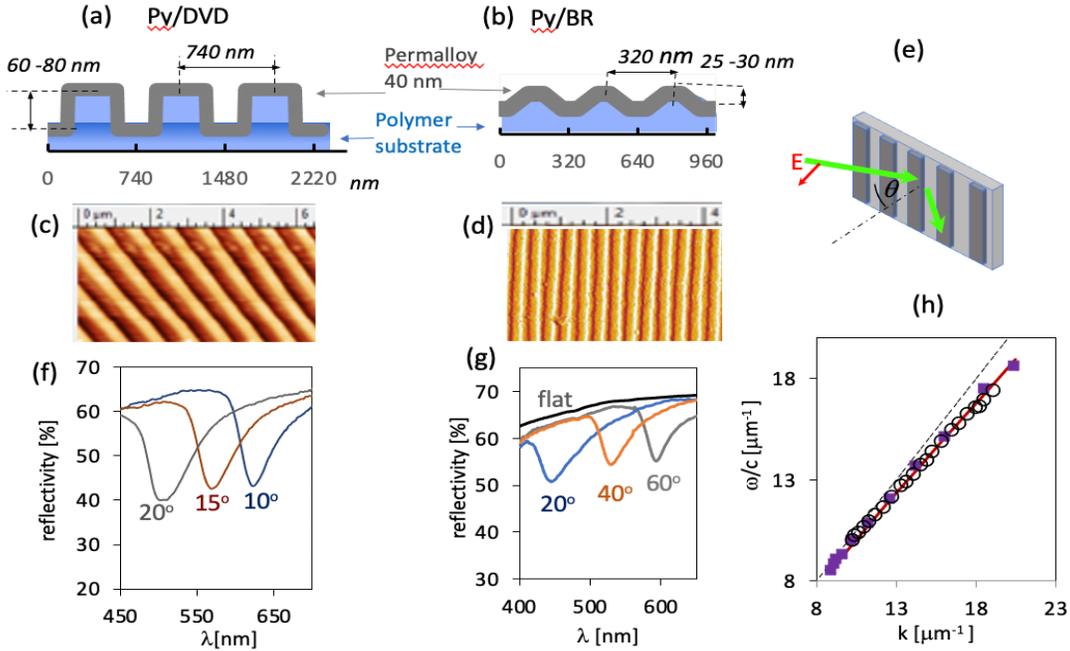

Fig. 1. (a, b) Schematics and (c,d) AFM images of Py/DVD and Py/BR structures respectively; (e) orientation of the sample in optical measurements; Reflectivity spectra at different incidence angles for (f) Py/DVD and (g) Py/BR structures; (h) dispersion relationship in Py/DVD (filled symbols) and Py/BR (open symbols), photon line (dashed trace) and theoretical estimation with Eqs 1,2 (red solid trace).

The photocurrent studies described below are performed with Py/DVD structures. The experimental setup is shown in Fig. 2 (a). The permalloy/DVD structure is prepared as a strip with a width of 3 mm and a length of 15 mm. The grooves are oriented perpendicularly to the long side. Two electric contacts are attached to the opposite ends of the strip. The sample is placed on a nonmagnetic stage and illuminated in the middle of the strip with the second harmonic of a Nd:YAG laser at 532 nm at p-polarization, with ~10 ns pulse duration and energy of ~ 0.15 mJ per pulse. The illumination spot covers the width of the sample. The voltage $U$ (Figs. 2 (b, c)) generated across the sample is measured with 2 GHz Tektronix Digital Oscilloscope with 50 $\Omega$ internal resistance. The magnetic field is supplied with an electromagnet. The direction of the magnetic field is parallel to the direction of the grooves.

In similarity with the profile-modulated gold and silver systems[18], strong enhancement of the photoinduced voltages is observed at the incidence angle corresponding to SPP excitation, Fig. 2 (b). The negative polarity of the signal at maximum corresponds to the drag of electrons in the direction of SPP propagation as expected from the momentum transfer consideration. However, the plasmon drag effect in this system presents a complicated picture, which could not be readily described with the simple EMHL approach[19]. The SPP-related voltages in permalloy are significant: $U/I$ (where $I$ is the illumination intensity) estimated for our experimental conditions is of $\sim 30$ mV/ (MW cm$^{-2}$). This value is significantly higher than typical $U/I$ in plasmonic gold and silver surfaces of 2 - 4 mV/ (MW cm$^{-2}$)[18,20]. An asymmetric Fano resonance-like shape of $U(\theta)$, Fig 2 (b), with the polarity switching back and forth in the range of SPP conditions is similar to the observations in Ag/DVD systems[32] where it is tentatively ascribed to coupling of propagating and localized plasmons. Alternatively, the switching to the opposite polarity may originate from a different mechanism (for example, photogalvanic effect[33]). In addition, a small non-zero signal is observed at normal incidence. It may be related to a slight asymmetry of geometry of an individual nanoscale feature[15] or a presence of a small non-zero magnetization which can contribute to photovoltages (in line with the findings described below). Acknowledging the complexity of the plasmon drag in our systems, in the present study we restrict the discussion only to the magnetic dependence of the photoinduced voltages.

In the experiment shown in Figs. 2 (c, d), we set the illumination angle in the range of the SPP resonance in order to achieve the maximal electric signal, and compare the signals at the magnetic field $H_0 = 85$ Oe directed down (-z direction) and up (+z). In Fig 2 (c), the SPP is excited in $y$ direction, and the photoinduced electric signals have negative polarity, which corresponds to the drag of electrons in $y$ direction. As one can see, the peak magnitudes of the photoinduced voltage are different for the opposite directions of the $H$ field. The difference between the signals $\Delta U = U (-H_0) - U(H_0)$ is shown in green. In Fig. 2 (d), the SPP is excited in the opposite direction, electrons drift in the -$y$ direction, but $\Delta U$ has the same polarity and almost the same amplitude as in the previous case. At SPP resonance conditions, $\Delta U$ is higher (up to 2-3 times) in magnitude than that that off resonance, see Fig. 2 (e). The polarity of $\Delta U$ always stays the same.

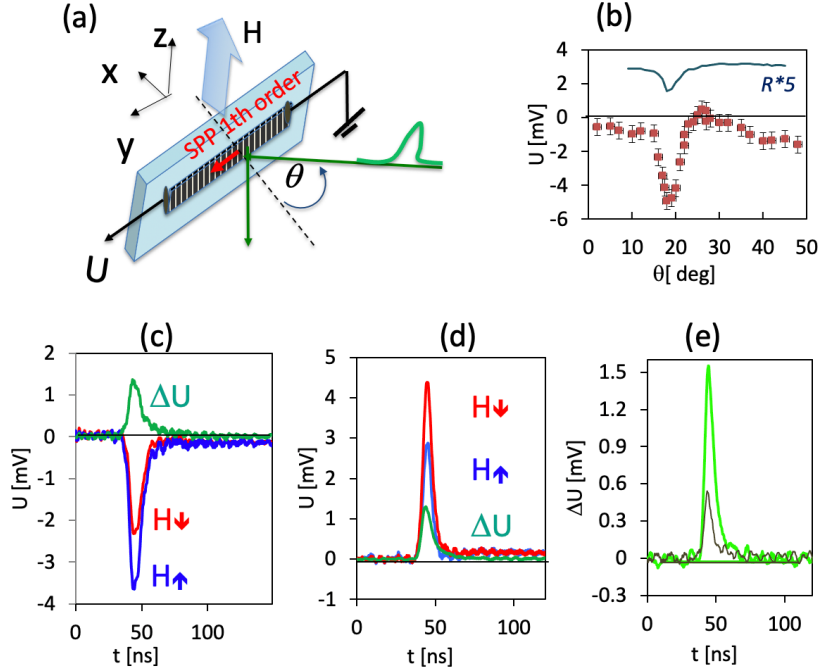

Fig. 2. (a) Experimental setup; (b) $U$ (red points) and reflectivity (blue solid trace) vs incidence angle $\theta$ at $H = 0$; (c) Typical photoinduced electric signals at the magnetic field $H = 85$ Oe directed down (red) and up (blue). Difference between them is shown in green; (d) Same at different direction of illumination; (e) $\Delta U$ at the SPP resonance conditions (light green) and off resonance (black).

In Fig. 3, the photoinduced voltage is recorded with respect to an applied external magnetic field $H$, which is slowly varied in the range between ±85 Oe. In Fig 3 (a), the field $H$ first increases from negative to positive values (step 1-2). The voltage of $\sim 3.7$ mV does not show any significant change in the magnitude until the applied magnetic field reaches to 45 Oe. A steep drop in $U$ from 3.7 mV to 2.4 mV is observed at 49 Oe. The voltage saturates at about 2.3 mV at a further increment of the magnetic field. Reversing the magnetic field sweep (step 2-3) does not affect the voltage until the field reaches negative values of ~ - 45 Oe. The voltage changes back to the original value 3.7 mV at ~ - 60 Oe with the further increase in $H$ in the –z direction, forming a full hysteresis loop.

In Fig. 3, the photoinduced voltage is recorded with respect to an applied external magnetic field $H$, which is slowly varied in the range between ±85 Oe. In Fig 3 (a), the field $H$ first increases from negative to positive values (step 1). The voltage of ~ 3.7 mV does not show any significant change in the magnitude until the applied magnetic field reaches to 45 Oe. A steep drop in $U$ from 3.7 mV to 2.4 mV is observed at 49 Oe. The voltage saturates at about 2.3 mV at a further increment of the magnetic field. Reversing the magnetic field sweep (step 2) does not affect the voltage until the field reaches negative values of ~ - 45 Oe. The voltage changes back to the original value 3.7 mV at ~ - 60 Oe with the further increase in $H$ in the –z direction, forming a full hysteresis loop.

In Fig. 3 (b), we test the effect of the field sweep when the SPP is excited in the opposite direction. Now we start from the zero field and vary the field in –z direction (step 1). Since the sample is previously exposed to negative fields, no switching is observed during this step. The voltage steeply changes from –5.4 mV to -6.1 mV only at the positive fields during the field sweep in +z direction (step 2) and stays at almost the same level during the next step 3. A similar hysteresis behavior is observed all Py/DVD samples, however the field where the switching in voltage occurs, slightly varies from sample to sample (of nominally the same composition), see an example in Fig. 3 (c). Note that the magnitudes of the switching fields are comparable with the anisotropy fields $H_a$ estimated from FMR characterization in our samples, indicating that the origin of the magnetic photocurrents is directly related to the magnetization of the sample. The slight variation in the behavior of the samples from different fabrication runs can be due to small imperfections or variations in the parameters of Py gratings ( Fig. 1 (a)), such as film thickness, modulation height or coverage of walls. Effect of such variations on the magnetic behavior of Py structures is the subject of the separate study which will be published elsewhere.

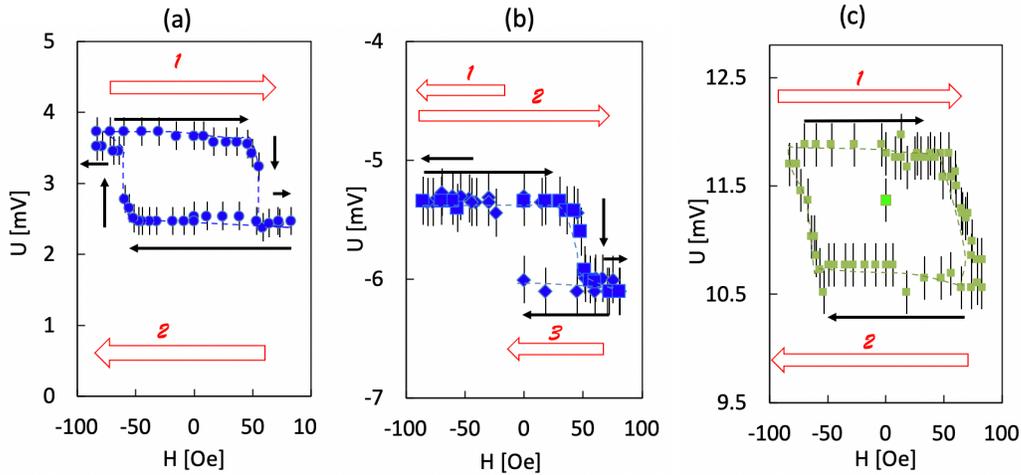

Fig 3. Hysteresis in photoinduced voltages with the variation of magnetic field. The field sweep direction is shown with the red arrows. Black arrows indicate the direction of the signal change. Dashed lines are a guide for the eye. Figs (a) and (b) correspond to different directions of illumination for the same sample. Data in (c) is obtained in a different sample. The data point shown in light green is obtained before introducing the field.

In principle, observed magnetodependence of the photoinduced voltage can be caused by a strong dependence of the electrical conductivity or optical permittivity on the magnetic field. However, the magnetoresistance of the bulk permalloy is quite low [33]. Our samples showed the relative change in the resistance of < 0.1 % as a response to the magnetic field $H$ = 3 kOe. We also tested a possible dependence of the SPP conditions on the external magnetic field, measuring the reflectivity at several angles at and around resonance conditions in different magnetic fields. We were not able to resolve any difference beyond the experimental error between the measurements without and with the field. We conclude that these factors do not play a significant role in magneto-dependent plasmon drag and can be omitted from consideration for now.

Summarizing our experimental observations, the magnetic part of the photoinduced voltage in profile-modulated permalloy films switches sign at the opposite directions of magnetization. It is significantly enhanced by SPPs, however shows the same polarity at the opposite directions of the SPP propagation.

The anomalous Nernst effect (ANE)[34-38] can be a possible mechanism. It predicts generation of the electric field in ferromagnetic materials in the presence of a thermal gradient $\nabla T$ and magnetization $M$, as

$$E_{ANE} = -N\mu_0 M \times \nabla T, \quad (3)$$

where $\mu_0$ is the permeability of a free space and $N$ is the Nernst coefficient. Let us assume that in our case the thermal gradient normal to the film plane can be created via light absorption along the penetration depth, $\delta$, see Fig. 4 (a), and the role of plasmonic resonance is only to achieve higher temperature gradient due to more efficient absorption.

In order to estimate $\nabla T$, consider the one-dimensional heat equation,

$$\frac{\partial u}{\partial t} = -\frac{\kappa}{c\rho}\frac{\partial^2 u}{\partial^2 x} + \frac{1}{c\rho}q, \quad (4)$$

where $u$ is the difference between the temperature at the particular depth, $T(x)$, and the ambient temperature $T_0$, $\kappa$ is the thermal conductivity, $c$ is the heat capacity, $\rho$ is the density. The function $q$ represents the density of effective "heat sources" associated with light absorption, and can be approximated as $q(x,t) = I(t)\frac{1-R}{\delta}e^{-\frac{x}{\delta}}$, where $I$ is the incident intensity, and $R$ is the reflectivity.

Eq. (4) is applied to the whole bilayer structure permalloy/substrate. We assume the standard values for permalloy, $c$=495 J/kg/K and $\rho$ =8700 kg/ m³, and use $\kappa$ = 20 W/(m*K)[38], and $\delta$ = 13 nm (as estimated from the data[26]). For the substrate (polycarbonate) we use $\kappa$ = 0.2 W/(m*K), $c$=1250 J/kg/K, and $\rho$ =1210 kg/ m³. As the boundary conditions, we neglect thermal exchange ($\frac{\partial u}{\partial x} = 0$) at both the outer surfaces and take into account the heat flow continuity at the permalloy-substrate interface ($x$= 40 nm), $\kappa_i\frac{\partial u_i}{\partial x} = \kappa_j\frac{\partial u_j}{\partial x}$, where $i$, $j$ denote permalloy or substrate material at the corresponding sides of the contact. Note that the penetration depth of light is 13 nm. During a short 10 ns pulse, only a thin surface layer having the permalloy film is efficiently heated while most of the substrate is at the ambient temperature. Assuming the thickness of the substrate to be 1 mm and parameters for polycarbonate as indicated above, the time between pulses, 0.1 s, is long enough for the permalloy film to cool down due to the heat conduction into the substrate and in the substrate. Estimating the laser induced heating (after a single pulse and a following relaxation), the total change in the temperature is ~ 0.8 mK if the substrate is completely thermally isolated from the environment. Since it is not the case (in particular, there is a good thermal contact between the substrate and the holder) we neglect the heating and assume the ambient temperature at the beginning of each pulse.

Eq. 4 is numerically solved, using the experimental shape of the pulse, $I(t)$, and absorbed pulse energy of 0.08 mJ. The temperature profiles $u$ $(x)$ are shown in Fig. 4 (b) at different moments of time during the laser pulse (beginning, peak and close to the end as indicated). The average temperature of the film grows toward the end of the pulse (by 25 K from the initial temperature), however the strongest slope is observed in the middle of the pulse. It reaches only about 0.35 K per 40 nm of the film thickness. Note that in permalloy, $N\mu_0 M$ = -9 nV/ K, and the observed voltage $U$ = 1.5 mV, Fig 2, would require $\nabla T = 4.2 * 10^7$ K/m, which is equivalent the temperature difference of ~ 2 K across the thickness of the film.

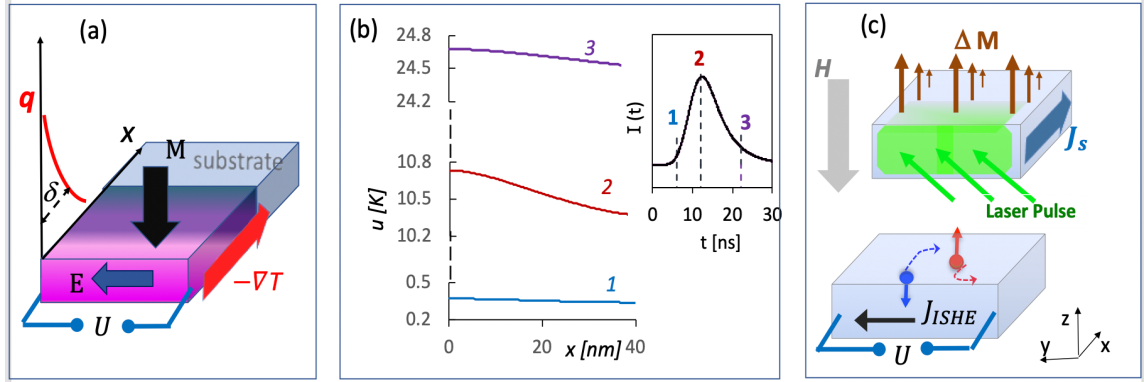

Fig. 4. (a) Geometry of ANE and distribution of effective heat sources q (x); (b) Solutions of Eq. 4 at the different moments of the laser pulse duration (as indicated at inset); (c) Possible ISHE mechanism in our system: Top: longitudinal ΔM associated with SAM transfer and spin current $J_s$; Bottom: Charge current $J_{ISHE}$.

Thus, the magneto-dependent plasmon drag effect observed in our systems is significantly stronger that what is predicted by a simple heating of the material. Without discarding this interpretation completely, one may consider other possible mechanisms, such as a direct effect of light on electron spin polarization, due to the presence of spin-orbital coupling. The spin angular momentum (SAM) transferred from plasmons to surface electrons causes a deviation of the electron spins from the equilibrium magnetization direction. As a result, the transverse spin component of the surface electrons exerts the exchange torque on the neighboring spins leading to spin current in $x$ direction. This mechanism is similar to that of "spin pumping" that arises under the condition of FMR excitation in bilayer structures (see, for example, the review article[40] and references therein).

The appearance of the transversal (rotating) spin component is accompanied by a decrease in longitudinal magnetization giving rise to a D.C component of the spin current (Fig. 4 (c)). Due to the inverse spin Hall effect (ISHE), this creates a D.C. charge current which is proportional to the vector product of the spin polarization and spin flow direction[41-42]. As a result, in our case, the corresponding potential difference arises along $y$ direction. When the sample magnetization is reversed, similar considerations yield the charge current in (-$y$) direction. $U(-H_0) – U(H_0)$ is positive which corresponds to the experimental observations. Naturally, this qualitative consideration is far from the final conclusion; further studies are planned.

Another alternative scenario can be an anomalous Hall effect (AHE)[43], which would generate charge current in the $y$ direction when electrons are pushed by light in $x$ direction. However, we believe that in our experiment this effect does not play a significant role, since it requires a unidirectional charge current across the film thickness during a relatively long period of time (at least, during the laser pulse), which is not expected in our geometry.

In conclusion, significant magnetic dependence of plasmon-enhanced photocurrents is observed in 1D profile-modulated permalloy films, manifesting the coupling of plasmonic, electric and magnetic effects. The mechanism of the phenomenon is not clear yet. Future experiments would provide more information on the origin of this coupling toward a better understanding of the energy, momentum and angular momentum transfer from light to matter in plasmonic systems and determine structures with the strongest responses. We believe these effects can find a variety of current and future applications ranging from compact magneto-optical devices, fast magnetization switching, nonreciprocal plasmonics, and plasmonics based electronics. In particular, the plasmon drag effect is promising for applications in plasmonic circuits [44, 45], optoelectronics with plasmonic elements[46] and plasmonic sensors with compact electric detection[32], since it provides the direct opportunity to monitor plasmonic excitations electrically. The possibility to control plasmon drag with a magnetic field brings an additional dimension to such applications.

**Acknowledgments**

The work was supported by NSF # 1830886, AFOSR FA9550-18-0417 and DoD #W911NF1810472 grants. Authors thank R. D. McMichael for the valuable discussion and V. V. Demidov for the help with FMR characterization,

**Data availability**

The data that support the results presented in the paper are available from the authors upon reasonable request.

**Author Contributions**

The first two authors contributed equally.

**Conflict of interest**

The authors declare that they have no conflict of interest.